\documentclass[twocolumn,pra,superscriptaddress]{revtex4}
\usepackage{bm}
\usepackage{graphicx}
\usepackage{amssymb}
\usepackage{mathrsfs}
\usepackage{amsmath}

\begin{document}

\title{Accelerating geometric quantum gates through non-cyclic evolution and shortcut to adiabaticity}

\author{Qing-Xian Lv}
 \affiliation{Guangdong Provincial Key Laboratory of Quantum Engineering and Quantum
Materials, School of Physics and Telecommunication Engineering,
South China Normal University, Guangzhou 510006, China}

\author{Zhen-Tao Liang}
 \affiliation{Guangdong Provincial Key Laboratory of Quantum Engineering and Quantum
Materials, School of Physics and Telecommunication Engineering,
South China Normal University, Guangzhou 510006, China}

\author{Hong-Zhi Liu}
 \affiliation{Guangdong Provincial Key Laboratory of Quantum Engineering and Quantum
Materials, School of Physics and Telecommunication Engineering,
South China Normal University, Guangzhou 510006, China}

\author{Jia-Hao Liang}
 \affiliation{Guangdong Provincial Key Laboratory of Quantum Engineering and Quantum
Materials, School of Physics and Telecommunication Engineering,
South China Normal University, Guangzhou 510006, China}

\author{Kai-Yu Liao}
\email{kaiyu.liao@foxmail.com}
 \affiliation{Guangdong Provincial Key Laboratory of Quantum Engineering and Quantum
Materials, School of Physics and Telecommunication Engineering,
South China Normal University, Guangzhou 510006, China}

\author{Yan-Xiong Du}
\email{duyanxiong@gmail.com}
 \affiliation{Guangdong Provincial Key Laboratory of Quantum Engineering and Quantum
Materials, School of Physics and Telecommunication Engineering,
South China Normal University, Guangzhou 510006, China}

\begin{abstract}
Fast and robust quantum gates is the cornerstone of fault-tolerance quantum computation. In this paper, we propose to achieve quantum gates based on non-cyclic geometric evolution. Dynamical phase during the evolution is cancelled by spin-echo process and the adiabatic control can be sped up through shortcut to adiabatic manner. Different from geometric gates based on cyclic evolution, the superiority of non-cyclic scheme is that the operation time is proportional to the rotation angle (but not the geometric phase) of quantum state. Therefore, the non-cyclic scheme becomes fairly fast in the case of quantum gates with small rotation angle which will be more insensitive to the decoherence and leakage to the states outside the computational basis. The proposed scheme is also robust against random noise due to the geometric characteristic of projective Hilbert space. Since the refined proposed scheme is fast and robust, it is an particularly suitable way to manipulate the physical systems with weak nonlinearity, such as superconucting systems.
\end{abstract}
\maketitle

\section{Introduction}
Geometric phase, which is accumulated during the cyclic evolution in the Hilbert space \cite{Berry1984,wilzeck1984,Aharonov1987,Aharonov1988}, is an important resource for quantum computation. The value of geometric phase depends on the solid angle enclosed by the evolution path and thus is insensitive to random noise along the path \cite{Sjoqvist2008,Sjoqvist2015,Sjoqvist2016}. There have been proposals of geometric quantum computation based on Abelian geometric phase \cite{zhusl2003}, non-Abelian geometric phase \cite{duan2001}, and non-adiabatic non-Abelian geometric phase \cite{Sjoqvist2012}. These proposals with geometric phases can not consolidate the superior of fast control and highly robust: the adiabatic manner is robust against parameters variation but running slowly; the non-adiabatic manner is fast but sensitive to the parameters variation. There are also schemes of realizing geometric quantum computation with shortcut to adiabaticity (STA) which accelerate adiabatic process through auxiliary Hamiltonian \cite{Liang2016,felix2018,Yu2018,Yan2019,Vepsalainen2019}, nevertheless, still should balance the weight between the speed and the energy cost \cite{campbell2017}.

One may conclude that the above proposals are all based on geometric phases induced in the cyclic evolution. A straight-forward result is that no matter how small the rotation (connecting the initial state and the desired state on the Bloch sphere) it is, the operation always needs a cyclic evolution of the cyclic states in the Hilbert space. As an ultimate case, the non-adiabatic non-Abelian geometric computation still needs a $\pi$ pulse area to realize arbitrary quantum gates \cite{Sjoqvist2012}. On the other hand, theories of geometric phase have been extended to the cases of non-cyclic Hamiltonian \cite{Aharonov1987} and non-cyclic evolution \cite{Samuel1988}. Proposals based on adiabatic perturbation theory have been came up to measure geometric/topologic information of Hilbert space through the non-cyclic evolution path \cite{Rigolin2008,Gritseva2012,Schroer2014,Roushan2014}. In this paper, we propose to realize quantum gates based on the non-cyclic geometric evolution. The dynamical phase during the adiabatic evolution can be cancelled through spin echo process and the final operation only depends on geometric parameters. The advantage of non-cyclic geometric evolution is that a smaller rotating angle corresponds to a shorter operation time, as compared with the cyclic manner. With a small enough rotation, the adiabatic non-cyclic control even operates faster than the one with non-adiabatic process. Therefore, such improvement on geometric quantum computation is more efficient than STA since STA can only accelerate adiabatic control by no more than 6 times with the same amplitude of control field \cite{Liang2016,du2016}. Furthermore, the adiabatic non-cyclic evolution can be incorporated with STA. The paper is organized as follow: In section II we introduce how to realize single qubit gate with non-cyclic geometric evolution. In section III the performance of proposed gate is tested. We conclude the paper in section IV.

\section{Realization of geometric gates with non-cyclic geometric evolution}

To illustrate how to realize quantum gates with non-cyclic evolution, considering a two-level system $\{|0\rangle, |1\rangle\}$ in a Xmon superconducting system which interacts with a microwave field as shown in Fig. 1a. Due to the weak non-linearity of the system, the microwave field may also couple the external states, i.e., the nearest level $|2\rangle$.The influence of $|2\rangle$ will be considered in the following section. The coupling Hamiltonian of $\{|0\rangle, |1\rangle\}$ is given by
\begin{equation}
H=\frac{\hbar}{2}\left(\begin{array}{cc}
\Delta &\Omega_{M}e^{-i\varphi}\\
\Omega_{M}e^{i\varphi}&-\Delta
\end{array}\right),
\end{equation}
where we have adopted the rotating wave approximation with $\Omega_M$ being the Rabi frequency of the control field and $\Delta$ being the detuning. The corresponding eigenstates are $|\lambda_-\rangle=\cos(\theta/2)|1\rangle-\sin(\theta/2)e^{-i\varphi}|0\rangle$, $|\lambda_+\rangle=\sin(\theta/2)e^{i\varphi}|1\rangle-\cos(\theta/2)|0\rangle$ and the eigenvalues are given by $E_{\pm}=\pm\hbar\Omega$, $\theta=\arctan\Omega_M/\Delta$. Here $\Omega=\sqrt{\Omega_M^2+\Delta^2}$.

\begin{figure}[ptb]
\begin{center}
\includegraphics[width=8cm]{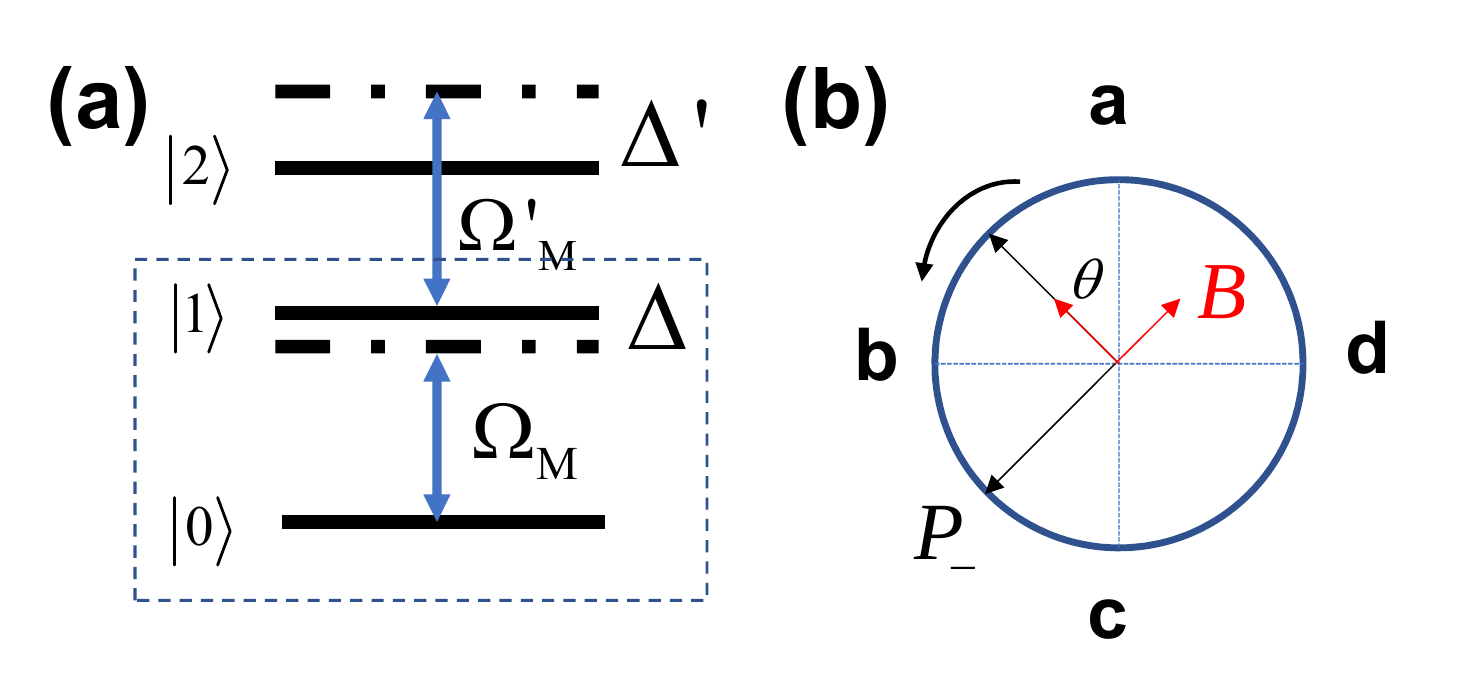}
\caption{ (color online). (a) Two-level system $\{|0\rangle, |1\rangle\}$ (surrounded by the blue-dashed rectangle box) interacts with a microwave field to realize single-qubit control in Xmon superconducting system. Due to the weak non-linearity of Xmon qubit, the microwave field may also couple the external states $|2\rangle$. (b) Diagrammatic sketch of realizing flip ($\sigma_x$ or $\sigma_y$) gate with non-cyclic evolution. $\mathbf{P_-}$ symbols the polarization vector of eigenstate $|\lambda_-\rangle$ which evolves along $a-b-c$. The effective magnetic field $\mathbf{B}$ that controls pseudo-spin evolves along the path $a-b-d-a$.}
\end{center}
\end{figure}

\begin{figure}[ptb]
\begin{center}
\includegraphics[width=7cm]{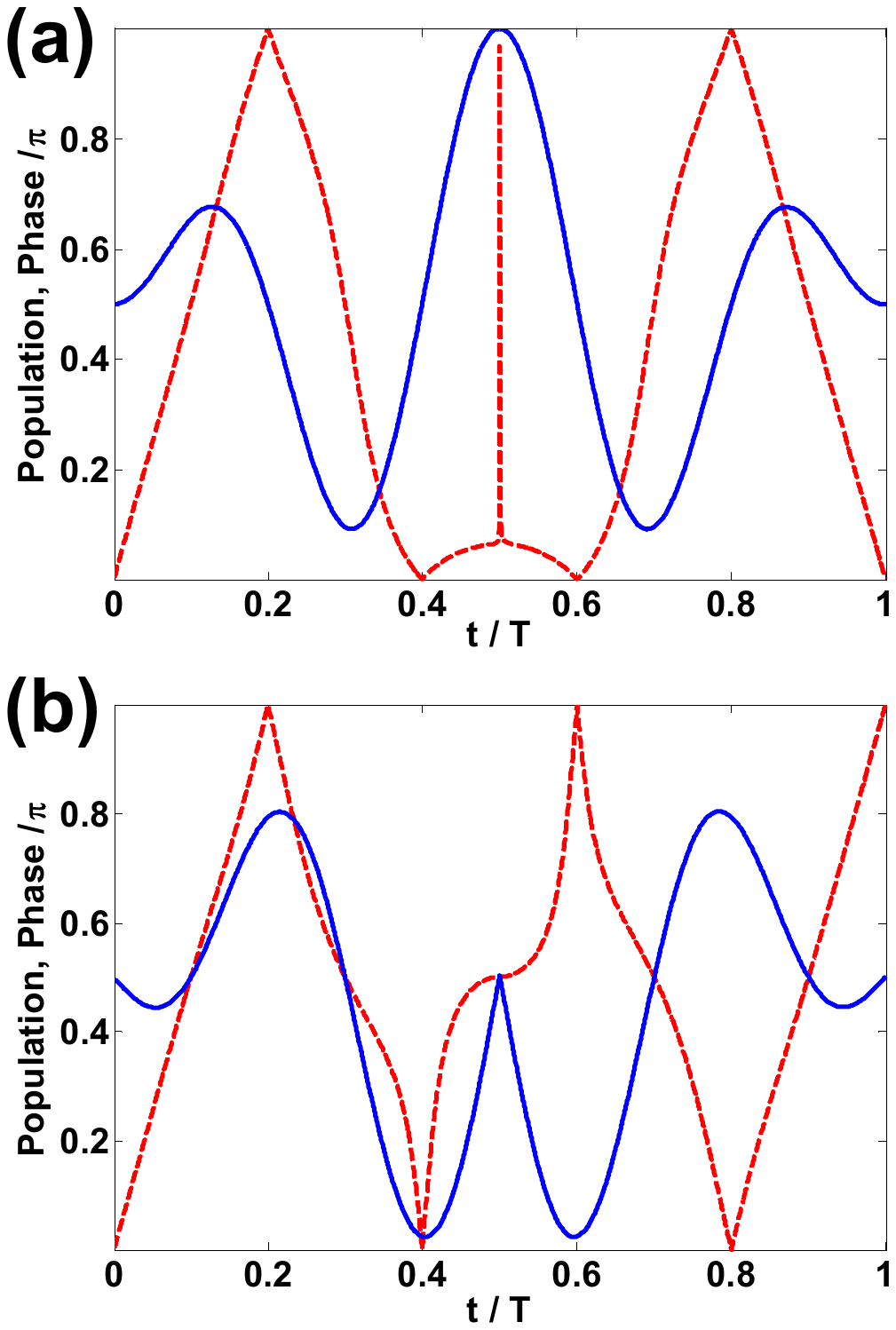}
\caption{ (color online). Population dynamics (blue-solid line) and the relative phases (red-dashed line) of geometric gates based on non-cyclic evolution. The initial state is $(1, 1)^{Ts}/\sqrt{2}$. (a) $\sigma_x$ gate. (b) $\sigma_y$ gate. The relative phase is in the unit of $\pi$.}
\end{center}
\end{figure}

The adiabatic control of the two-level system can be realized by driving $\theta$ slowly enough to satisfy the adiabatic condition ($\dot{\theta}\ll(E_+-E_-)/\hbar$). With the control parameters $\theta:0\rightarrow\theta_1, \varphi$, any initial state $|\psi(0)\rangle=\alpha|0\rangle+\beta e^{i\delta}|1\rangle$ under adiabatic driving will induce the bare states evolving along the eigenstates: $|0\rangle\rightarrow|\lambda_+\rangle$, $|1\rangle\rightarrow|\lambda_-\rangle$. The final state is given by $|\psi(T)\rangle=\alpha e^{-iA_+}|\lambda_+\rangle+\beta e^{i\delta}e^{-iA_-}|\lambda_-\rangle$ \cite{Du2014}. $A_\pm$ is the dynamical phases. The geometric phases of the eigenstates are omitted here since it is vanishing when the eigenstates evolve along the longitude line. Evolution operator $U_1=U(\theta:0\rightarrow\theta_1, \varphi)$ that connects the initial and the final state is given by
\begin{equation}
U_1=\left(\begin{array}{cc}
\cos(\frac{\theta_1}{2}) & -\sin(\frac{\theta_1}{2})e^{-i\varphi}\\
\sin(\frac{\theta_1}{2})e^{i\varphi} & \cos(\frac{\theta_1}{2})
\end{array}\right)\left(\begin{array}{cc}
e^{-iA_+} & 0\\
0 & e^{-iA_-}
\end{array}\right).
\end{equation}
The evolution operator $U_2=U(\theta:\theta_1\rightarrow 0, \varphi)$ is easily obtained by checking $U_2U_1=\left(\begin{array}{cc}
e^{-2iA_+} & 0\\
0 & e^{-2iA_-}
\end{array}\right)$ which is derived as
\begin{equation}
U_2=\left(\begin{array}{cc}
e^{-iA_+} & 0\\
0 & e^{-iA_-}
\end{array}\right)\left(\begin{array}{cc}
\cos(\frac{\theta_1}{2}) & \sin(\frac{\theta_1}{2})e^{-i\varphi}\\
-\sin(\frac{\theta_1}{2})e^{i\varphi} & \cos(\frac{\theta_1}{2})
\end{array}\right).
\end{equation}
The union operation of $U_{21}=U(\theta:\theta_1\rightarrow 0, \varphi+\pi)U(\theta:0\rightarrow\theta_1, \varphi)$ is given by
\begin{equation}
U_{21}=\left(\begin{array}{cc}
\cos\theta_1e^{-2iA_+} & -\sin\theta_1e^{-i\varphi}\\
\sin\theta_1e^{i\varphi} & \cos\theta_1e^{-2iA_-}
\end{array}\right).
\end{equation}
One may see that the dynamical phase is ignored when $\theta_1=\pi/2$. It is in fact the result of spin-echo effect. By abstracting a common phase, $U_{21}$ becomes a $\sigma_x$ operation when $\varphi=\pi/2$ and is a $\sigma_y$ gate when $\varphi=0$. The trajectory of $|\lambda_-\rangle$ (symbolled by the polarization vector $\mathbf{P}_-=\langle\lambda_-|\mathbf{\sigma}|\lambda_-\rangle$, $\mathbf{\sigma}$ are the Pauli Matrixes) is shown in Fig. 1b which evolves along the path $a-b-c$. It is obvious that the adiabatic eigen basis has not evolved cyclically which is different from the previous proposals \cite{zhusl2003,duan2001,Sjoqvist2012}. Through parameterizing the effective control field acting on the quantum states by $\mathbf{B}=(\sin\theta\sin\varphi, \sin\theta\cos\varphi, \cos\theta)$, the trajectory is shown to be evolveing along the path $a-b-d-a$ as plotted in Fig. 1b.

Following similar calculation as above, any arbitrary rotation can be obtained by
\begin{eqnarray}
U_r&=&U(\theta:\pi-\theta_1\rightarrow \pi-2\theta_1, \varphi+\pi)U(\theta:0\rightarrow\theta_1, \varphi)\notag\\
&=&\left(\begin{array}{cc}
\cos\theta_1 & -\sin\theta_1e^{-i\varphi}\\
\sin\theta_1e^{i\varphi} & \cos\theta_1
\end{array}\right),
\end{eqnarray}
where $\theta_1\in[0,\pi/2)$. One can find that the operation is purely depended on the geometric parameters as $U_r=U_r(\theta_1, \varphi)$. Here both the evolution path of adiabatic eigen basis and the control parameters in operation Eq.(5) is non-cyclical. Also, the dynamical phase is cancelled through spin-echo process and the geometric phase of the eigen basis is zero along the longitude line. The geometric control on arbitrary states here origins from the parallel transport of the eigen basis in the projective Hilbert space \cite{Grigorenko}. Note that our proposal is somewhat similar to the proposal in \cite{zheng2005,zheng2009} which performs non-cyclic manipulation on dark state. However, such proposal based on dark states can not be used to realize universal single-qubit gates. As a generalization, we perform manipulation on two eigen basises to achieve universal single qubit gates after cancelling the dynamical phase. Besides, proposal \cite{zheng2005} can be treated as a two-qubit gate version of our work. Actually, our proposal can be used to obtain two-qubit controlled-phase gate in Xmon superconducting system which will polish up the previous adiabatic frequency-sweeping version \cite{Barends2014}.

\section{Performance of the geometric gates}

In the following the process of achieving non-cyclic single-qubit gates with suitable waveforms is presented. We set the Rabi frequency and the detuning to be
\begin{equation}
\Omega_M(t)=\Omega\sin(\frac{\pi t}{T})e^{i\varphi}, \Delta(t)=\Omega\cos(\frac{\pi t}{T}),
\end{equation}
with the driving parameter $\theta=\pi t/T$. To realize a flip gate, we divide the process into two stages: at the first stage $t:0\rightarrow T/2, \varphi=\varphi_0$; at the second stage $t:T/2\rightarrow T, \varphi=\varphi_0+\pi$. When $\varphi_0=\pi/2$, we achieve a $\sigma_x$ gate (rotating along x axis by $\pi$ angle, symbolled by $R_x(\pi)$); when $\varphi_0=0$, we achieve a $\sigma_y$ gate (rotating along x axis by $\pi$ angle, symbolled by $R_y(\pi)$). The adiabatic evolution can be accelerated by STA through applying an auxiliary control Hamiltonian which is given by $H_{cd}=i\dot{\theta}(e^{-i\varphi}|0\rangle\langle1|+e^{i\varphi}|1\rangle\langle0|)$ \cite{Chen2010}. It is easy to check the STA Hamiltonian won't contribute dynamical phase. The population dynamics (blue-solid line) and the relative phases (red-dashed line) under the non-cyclic geometric control with STA is shown in Fig. 2. We adopt $\Omega T=5\pi$ to confirm the auxiliary field to be much weaker than the original field. The relative phases of states are determined by $\chi=\arccos(\langle\sigma_x\rangle/\sqrt{\langle\sigma_x\rangle^2+\langle\sigma_y\rangle^2})$ where $\langle\bullet\rangle$ is the average value of the Pauli Matrix. Fig. 2a is the result of $\sigma_x$ gate and Fig. 2b is the result of $\sigma_y$ gate. To remove the phase singularity at the poles, we choose the initial state as $|\psi(0)\rangle=(1, 1)^{Ts}/\sqrt{2}$, $Ts$ is the transposition. The acquired relative phases after the evolution are accord with the theoretical prediction as shown in Fig. 2.

In Fig. 3a we test the robustness of the proposed $\sigma_x$ gate against the variation of Rabi frequency. The variation is introduced by $\Omega'=\eta\Omega$. The fidelity is defined as $F=\sqrt{Tr(\rho_{ideal}\rho)}$, where $\rho$ is the density matrix of the actual final state and $\rho_{ideal}$ is the ideal one, $Tr$ denotes the trace of density matrix. As shown by the red- dashed line in Fig. 3a, the fidelity of proposed gate keeps high in a range of $20\%$ variation and is higher than $0.99$ at a range of $10\%$ variation. As a comparison, the $\sigma_x$ gate based on Rabi oscillation with the same amplitude of control field drops quickly if the Rabi frequency changes. We also compare the proposal scheme with the composite pulses \cite{Rong2015}. The black-solid line is the numerical result of the refocusing pulses which is achieved by $R_y(\pi/2)R_x(\pi)R_y(\pi/2)$ \cite{Vandersypen2005}. The robustness of the non-cyclic scheme is comparable with the composite pulses method and thus without the need of strengthen by composite pulses, unlike the case in non-adiabatic scheme \cite{chen2018}. A comparison on detuning variation has a similar result as the case of Rabi frequency. Therefore, the proposal is robust against the systematic errors.

\begin{figure*}[ptb]
\begin{center}
\includegraphics[width=12cm]{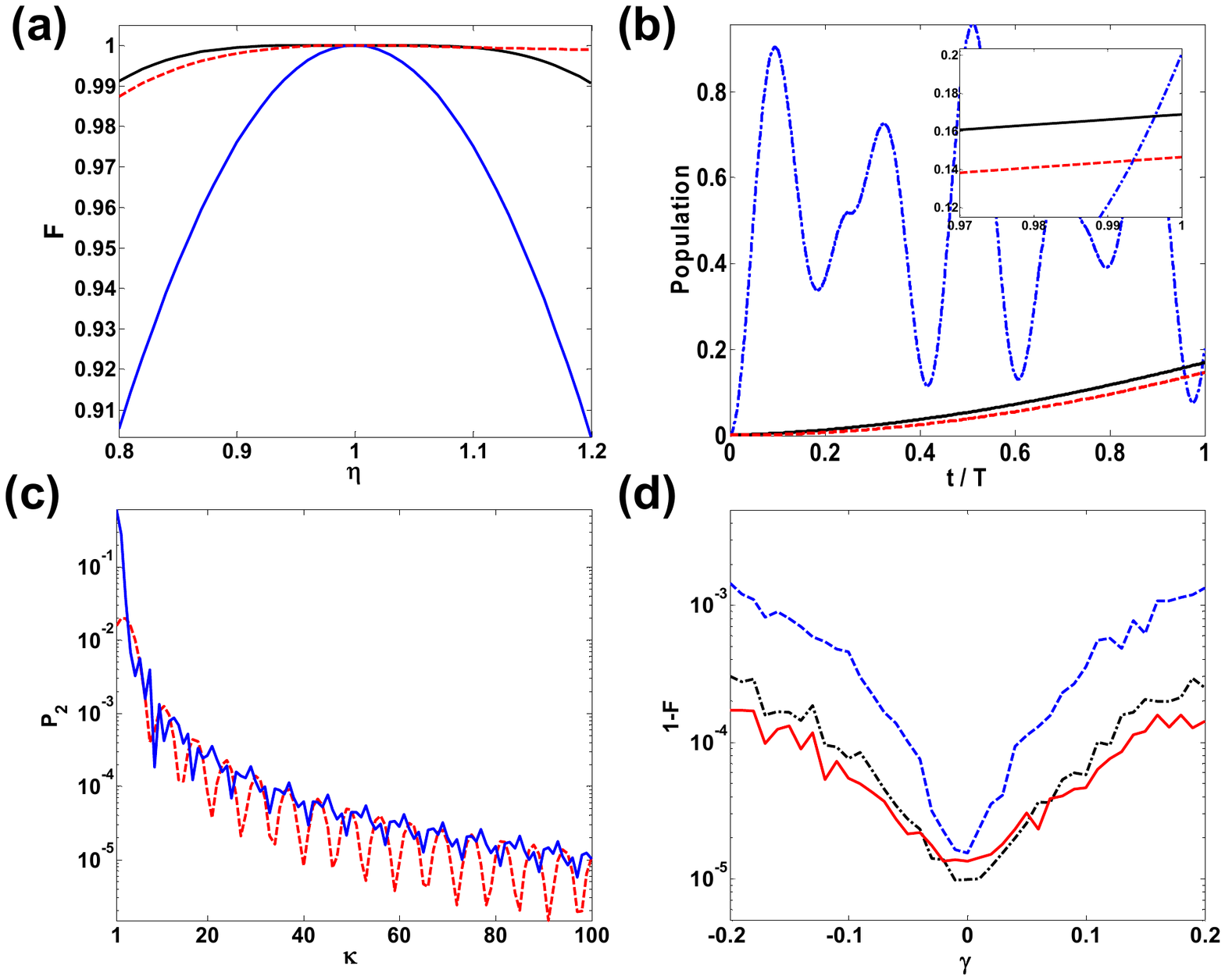}
\caption{ (color online). (a) Robustness of the proposed $\sigma_x$ gate against the variation of Rabi frequency. The variation is introduced by $\Omega'=\eta\Omega$. The plot of $\sigma_x$ gate based on Rabi oscillation and composite pulses are used as comparison. Red-dashed line: non-cyclic scheme. Blue- solid line: Rabi oscillation. Black-solid line: composite pulse. The proposed scheme has been sped up by STA and the evolution time is 5 times of Rabi oscillation. It can be seen that the robustness of the non-cyclic scheme is much stronger than the Rabi oscillation and is comparable with the composite pulse. (b) Comparison of the $\pi/8$ rotation gate (along y axis) realized by non-cyclic/cyclic scheme under decay process. $\pi/8$ rotation gate in the cyclic scheme is achieved by SGQG with three stages. The red-dashed line is the ideal case of non-cyclic scheme while the black-solid line is the one include a decay rate of $\Gamma=2\pi\times20$ kHz. The blue dotted-dashed line is the result of SGQG with the same decay rate. As shown in the inset, The cyclic scheme will suffer more from the decay since its operation time is 8 times longer than the non-cyclic scheme. The evolution time has been normalized for the convenience of comparison. (c) Comparison of leakage between the proposed scheme (red-dashed line) and SGQG (blue-solid line). (d) Comparison of the $\sigma_x$ gate realized by non-cyclic/cyclic scheme process against random noise. The random noise has a zero mean value over the evolution with an amplitude of $\gamma\Omega$. The red-solid line and the black dotted-dashed line are the numerical result of non-cyclic scheme and the cyclic scheme with same frequency of noise, respectively. The blue-dashed line is the result of cyclic scheme with a 2 times lower frequency. The robustness of non-cyclic scheme is as good as the cyclic scheme with noise of the same frequency. The data has been averaged by 50 times.}
\end{center}
\end{figure*}

We compare the effect of decoherence on geometric quantum gate with scheme of cyclic/non-cyclic evolution in Fig. 3b. The geometric quantum gate with cyclic evolution scheme is realized by superadiabatic geometric quantum gate (SGQG) which consists by three stages \cite{Liang2016}. The rotation angle of quantum gate with SGQG is depend on the solid angle surrounded by the ``orange slice". Thus no matter how small the rotation angle it is, the evolution time is always the same. The case is different in the non-cyclic evolution. As can be seen in Eq.(5) and (6), the rotation angle in our scheme is depend on $\theta$ which is proportional to $t$. A decrease in the rotation angle will shorten the evolution period linearly. Furthermore, the non-cyclic scheme will operate even faster if one raise the amplitude of control field as the same level with the cyclic scheme.  In the comparison in Fig. 3b, we use a $\pi/8$ rotation gate along $y$ axis $R_y(\pi/8)=e^{i\pi/8\sigma_y}$. Gate realized by SGQG needs a $\pi/8$ azimuthal angle with $2T$ evolution time (we use spin-echo in a single closed loop) while the one with non-cyclic evolution only needs $T/4$ (two stages with $T/8$ evolution time) which is 8 times faster than SGQG. We simulate the decoherence through master equation as \cite{Lindblad1976,Johansson2012}
\begin{equation}
\dot{\rho}=-i[H,\rho]+2L\rho L^\dagger-L^\dagger L\rho-\rho L^\dagger L,
\end{equation}
where $L=\sqrt{\Gamma}|0\rangle\langle 1|$ and $\Gamma$ is the decay rate. In the simulation we set $\Gamma=2\pi\times20 $ kHz, the Rabi frequency $\Omega=2\pi\times5$ MHz with the waveform (6), and $\Omega T=5\pi$ \cite{Yan2019}. Both two schemes are accelerated by STA. We have normalized the evolution time for the convenience of comparison in the plot. The red-dashed line is the population dynamics of non-cyclic scheme in the ideal case and the black-solid line is the one which introduced decay process. The blue dotted-dashed line is the dynamics of cyclic scheme. As shown in the inset of Fig. 3b, the final population of the proposed non-cyclic scheme is more close to the ideal case than SGQG since it is faster to be more insensitive to the decay. Then we further test the performance of cyclic/non-cyclic scheme in a weak non-linear system, through analyzing the population leakage to state $|2\rangle$, $P_2$ . The interacting Hamiltonian in the basis $\{|0\rangle, |1\rangle, |2\rangle\}$ is derived as \cite{Motzoi2009}
\begin{equation}
H_3=\frac{\hbar}{2}\left(\begin{array}{ccc}
0 &(\Omega_{M}-i\dot{\theta})e^{-i\varphi} & 0\\
(\Omega_{M}+i\dot{\theta})e^{i\varphi}& -2\Delta  & (\Omega'_{M}-i\dot{\theta'})e^{-i\varphi}\\
0 & (\Omega'_{M}-i\dot{\theta'})e^{i\varphi} & -2\Delta'
\end{array}\right),
\end{equation}
where $i\dot{\theta} (i\dot{\theta'})$ is the counter diabatic term and $\Omega'_{M}=h\Omega_{M}$ and $\dot{\theta'}=h\dot{\theta}$ have considered the different coupling between levels, $\Delta'=\Delta-\Delta_0$ with $\Delta_0$ is the energy difference between $|0\rangle,|1\rangle$ and $|1\rangle,|2\rangle$. As shown in Fig. 3c, the blue-solid line is the leakage after the $\pi/8$ rotation gate realized by SGQG while the red-dashed line is the one of non-cyclic scheme. In the simulation we have chosen $\Omega=2\pi\times5$ MHz, $\Omega T=5\pi$ and $h=0.8$. It can be seen that in a large range of detuning $\Delta_0$, the non-cyclic scheme has much smaller leakage than SGQG (up to 10 times) where we have parameterized $\Delta_0=\kappa\Omega$. SGQG works worse than the non-cyclic scheme in a small rotation angle is mainly due to it pumps a large fraction of population to $|1\rangle$ during the evolution which is unnecessary, as shown in Fig. 3b.

In Fig. 3d, we test the robustness of the proposed $\sigma_x$ gate against random noise. The random noise is artificially introduced to the amplitude of Rabi frequency by $\Omega_r=(1+\gamma)\Omega$ with $\gamma\in(-0.2,0.2)$. The random noise will cause dynamical dephasing since the dynamical phase is only cancelled at the end but is accumulated during the evolution \cite{Berger2013,zhu2005,Chiara2003}. The red-solid line is the numerical infidelity of non-cyclic scheme. We insert 1000 points with a vanishing mean value into the Rabi frequency during the simulation. As comparison, the plot of SGQG (blue-dashed line) is also shown with 1000 points inserted during the evolution. The performance of SGQG is significantly worse than the non-cyclic scheme since the frequency of noise is 2 time lower than the one of non-cyclic scheme (the evolution time of SGQG is 2 times longer than the non-cyclic one). However, if we increase the frequency of noise on SGQG by 2 times, the performance of cyclic scheme is as good as the non-cyclic scheme. Therefore, the robustness of the non-cyclic scheme against the random noise is comparable with the geometric control with cyclic geometric phase.

\section{conclusion}

In conclusion, we have shown that our geometric control is actually the effect of parallel transport on the projective Hilbert space since the geometric phase is vanishing and the dynamical phase has been cancelled \cite{zheng2005}. The abandon of the cyclic evolution condition allows us to operate the state much faster. The robustness of our scheme relies on the adiabaticity and the geometric characteristic. It is also possible to generalize the cyclic manipulation to the non-Abelian case \cite{Unanyan1999}. Therefore, the non-cyclic geometric control will provide a new way for quantum control and quantum computation.

\section*{Acknowledgments}

This work was supported by the NSF of China (Grants No. 11704131, 61875060 and
11125417), the Major Research Plan of the NSF of China (Grant No.
91121023), the  NFRPC (Grant No. 2011CB922104), the FOYTHEG (Grant
No. Yq2013050), the PRNPGZ (Grant No. 2014010), and the PCSIRT
(Grant No. IRT1243).

\begin{figure}[ptb]
\begin{center}
\includegraphics[width=9cm]{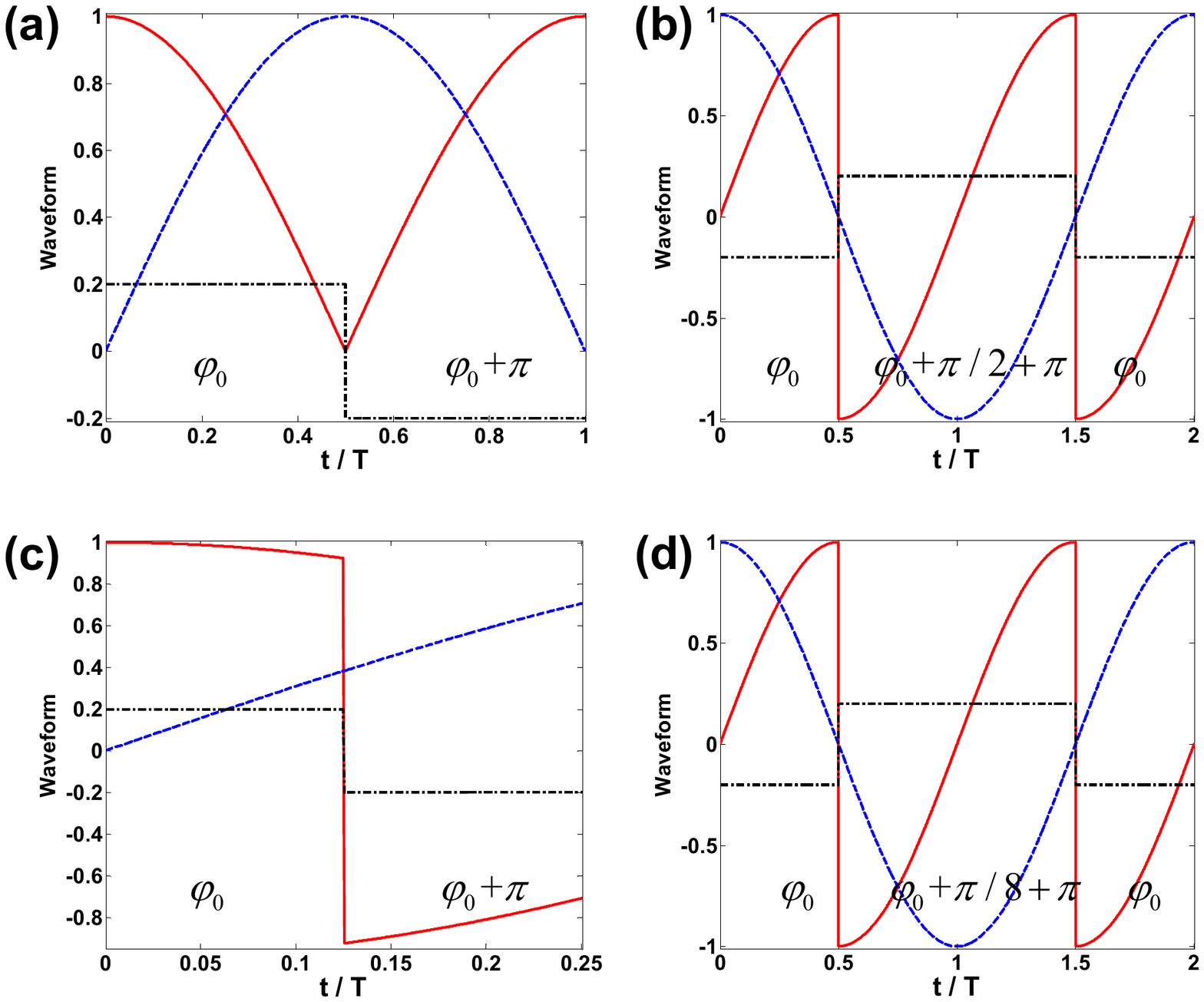}
\caption{ (color online). Waveforms of Rabi frequencies and detuning to realize geometric quantum gates. Red solid lines: detuning. Blue-dashed lines: Rabi frequency. Black dotted-dashed lines: auxiliary Rabi frequency. (a) $\sigma_x$ gate with non-cyclic scheme. (b) $\sigma_x$ gate with cyclic scheme. (c) $\pi/8$ rotation gate with non-cyclic scheme. (d) $\pi/8$ rotation gate with cyclic scheme. All Rabi frequencies the plots are in the unit of $\Omega$ and $t$ is in the unit of $T$. The relative phases of Rabi frequencies are also shown in the plots. As can be seen that non-cyclic scheme has the virtue of operating in a small rotation angle.}
\end{center}
\end{figure}

\section*{Appendix A: Derivation of arbitrary rotation gate with non-cyclic evoltuion}
In the following, we describe how to obtain arbitrary rotation gate (5) with the method of evolution operator. By using the transitivity of evolution operator $U(\theta:0\rightarrow\theta_2,\varphi)=U(\theta:\theta_1\rightarrow\theta_2,\varphi)U(\theta:0\rightarrow\theta_1,\varphi)$, evolution operator that describes the transportation from $\theta_1$ to $\theta_2$ along adiabatic evolution is derived as
\begin{widetext}
\begin{eqnarray}
&&U(\theta:\theta_1\rightarrow\theta_2,\varphi)\\\notag
&&=\left(\begin{array}{cc}
\sin(\frac{\theta_2}{2})\sin(\frac{\theta_1}{2})e^{-iA_-}+\cos(\frac{\theta_2}{2})\cos(\frac{\theta_1}{2})e^{-iA_+} & -\sin(\frac{\theta_2}{2})\cos(\frac{\theta_1}{2})e^{-i\varphi}e^{-iA_-}+\cos(\frac{\theta_2}{2})\sin(\frac{\theta_1}{2})e^{-i\varphi}e^{-iA_+}\\
-\cos(\frac{\theta_2}{2})\sin(\frac{\theta_1}{2})e^{i\varphi}e^{-iA_-}+\sin(\frac{\theta_2}{2})\cos(\frac{\theta_1}{2})e^{i\varphi}e^{-iA_+} & \cos(\frac{\theta_2}{2})\cos(\frac{\theta_1}{2})e^{-iA_-}+\sin(\frac{\theta_2}{2})\sin(\frac{\theta_1}{2})e^{-iA_+}
\end{array}\right).
\end{eqnarray}
\end{widetext}
Therefore, through cancelling the dynamical phase, one can obtain arbitrary rotation gate by $U_r=U(\theta:\pi-\theta_1\rightarrow \pi-2\theta_1, \varphi+\pi)U(\theta:0\rightarrow\theta_1, \varphi)$.

\section*{Appendix B: Waveforms of geometric quantum gates}
Here we plot the waveforms to realize $\sigma_x$ gate and $\pi/8$ rotation gate with non-cyclic/cyclic scheme. As shown in Fig. 4, the evolution time to achieve $\sigma_x$ gate with non-cyclic scheme (Fig. 4a) is 2 times shorter than the one of cyclic scheme (Fig. 4b), while the evolution time to achieve $\pi/8$ rotation gate with non-cyclic scheme (Fig. 4c) is 4 times shorter than $\sigma_x$ gate with non-cyclic scheme, and is 8 times shorter than $\pi/8$ rotation gate with cyclic scheme (Fig. 4d). Therefore, the non-cyclic scheme is faster than the cyclic scheme, and has the virtue of operating in a small rotation angle.

\end{document}